\documentclass{aa}
\usepackage{graphics}
\usepackage{natbib}
\bibpunct{(}{)}{;}{a}{}{,}

\newcommand{\Lya}{Ly$\alpha$}
\newcommand{\Lyb}{Ly$\beta$}
\newcommand{\cq}{\ion{C}{iv}}
\newcommand{\ctre}{\ion{C}{iii}}
\newcommand{\cdue}{\ion{C}{ii}}

\newcommand{\siq}{\ion{Si}{iv}}
\newcommand{\sitre}{\ion{Si}{iii}}
\newcommand{\sidue}{\ion{Si}{ii}}
\newcommand{\mgd}{\ion{Mg}{ii}}

\newcommand{\nc}{\ion{N}{v}}
\newcommand{\ntre}{\ion{N}{iii}}
\newcommand{\huno}{\ion{H}{i}}
\newcommand{\osei}{\ion{O}{vi}}
\newcommand{\aldue}{\ion{Al}{ii}}
\newcommand{\altre}{\ion{Al}{iii}}

\newcommand{\cm}{cm$^{-2}$}
\newcommand{\kms}{km s$^{-1}$}
\newcommand{\lsim}{\raisebox{-5pt}{$\;\stackrel{\textstyle <}{\sim}\;$}}
\newcommand{\gsim}{\raisebox{-5pt}{$\;\stackrel{\textstyle >}{\sim}\;$}}

\begin{document}


   \title{Inhomogeneous metal enrichment at $z\sim 1.9$: 
          the Lyman limit systems in the spectrum of the
          HDF-S quasar\thanks{Based on material collected 
          during commissioning time of UVES, the UV and
          Visible Echelle Spectrograph, mounted on the
          Kueyen ESO telescope operated on Cerro Paranal
          (Chile) and with the NASA/ESA {\sl Hubble Space 
   Telescope}, obtained at the Space Telescope Science Institute,
   which is operated by the Association of Universities of Research in
   Astronomy, Inc. under NASA contract NAS 5-26555.}}

   \author{ Valentina D'Odorico\inst{1} \and Patrick Petitjean\inst{1,2}}

   \institute{Institut d'Astrophysique de Paris, 98bis Boulevard
        Arago, F-75014 Paris  
     \and
        UA CNRS 173 - DAEC, Observatoire de Paris Meudon, F-92195
        Meudon Principal Cedex, France}

   \offprints{V. D'Odorico}
%
 
   \date{Received; accepted }

    \titlerunning{Inhomogeneous metal enrichment at
                  $z\sim 1.9$}
 
    \authorrunning{V. D'Odorico \& P. Petitjean}

\abstract{
%
We present a detailed analysis of three metal absorption
systems observed in the spectrum of the HDF-South quasar 
J2233-606 ($z_{\rm em}=2.238$), taking advantage of 
new VLT-UVES high resolution data ($R=45000$, S/N$=40-60$, 
$\lambda\lambda\,3050-10000$ \AA). 
%
Three main components, spanning about 300 \kms, can
be individuated in the Lyman limit system at $z\sim
1.92$. They show a surprisingly large variation in
metallicities, respectively $\sim 1/500$, $1/8$ and
$1/100$ solar. The large value found for the second
component at $z\simeq1.9259$, suggests that
the line of sight crosses a star-forming region. 
In addition, there is a definite correlation between
velocity position and ionisation state in this component, 
which we interpret as a possible 
signature of an expanding \ion{H}{ii} region.
The systems at $z\sim 1.94$ and $z\sim 1.87$ have also 
high metallicity, $\sim 1/4$ and $1/3$ solar. 
We find that photoionisation and collisional ionisation
are equal alternatives to explain the high excitation
phase revealed by \osei\ absorption, seen in these two
systems.  
From the width of the \siq, \cq, \sitre\ and \ctre\
lines in the system at $z\sim 1.87$, we can estimate 
the temperature of the gas to be $\log T \lsim 4.7$,
excluding collisional ionisation.
Finally, we compute the \siq/\cq\ ratio for all Voigt
profile components in a sample of $\log N$(\cq)$ \gsim
14$ systems at $z \lsim 2$. 
The values show a dispersion of more than an order of
magnitude and most of them are much larger than
what is observed for weaker systems. 
This is probably an indication that high column
density systems preferably originate in galactic halos
and are mostly influenced by local ionising sources. 
}

   \maketitle

%

\section{Introduction}



High \huno\ column density systems 
($N$(\huno)~$\gsim$~10$^{16}$ cm$^{-2}$)
and especially Lyman limit systems 
($N$(\huno)~$\gsim$~10$^{17}$ cm$^{-2}$, hereafter LLS)
arise in dense environments such as halos of large galaxies
or the densest regions of filaments linking the galaxies
\citep[e.g.][]{katz96,gardner97}. 
Indeed, at redshifts $z\lsim 1$, galaxies associated with  
LLS (detected by \mgd\ absorption with equivalent 
width $w_{\rm r}$~$>$~0.3~\AA) are routinely 
identified revealing the presence of gaseous 
halos with radius larger than $\approx 40\ h^{-1}$ kpc    
\citep{berg91,steidel94,church96,guill,church00a,church00b}. 

Although the ionisation corrections are large in LLS, the
large number of associated metal lines can be used to
constrain ionisation 
models and to estimate metallicity, ionisation state and
abundances 
\citep[e.g.][]{ppj92,berg94,kt99,kohler99,pb99,cp00}.  

In this paper, we present the study of three metal systems at
redshifts $z_{\rm abs} \sim $ 1.87, 1.92, and 1.94 seen
in the spectrum of the HDF-South quasar J2233-606 
\cite[$z_{\rm em} = 2.238$, $B \simeq
17.5$; see][]{sav98,sav99}.  
The systems are associated with a
Lyman limit break observed at $\sim 2683$ \AA\ \citep{sdw98}.
\citet[][ PB99]{pb99} investigated the two systems at $z
\sim 1.92$ and $z\sim 1.94$ in detail. They 
derived $\log N$(\huno) $= 17.15 \pm 0.02$ and $16.33 \pm 0.04$
at $z \sim 1.92$ and 1.94 respectively.
They found the systems have solar abundance pattern, but 
significantly different metallicites: $\approx 1/50$
solar and less than $\approx 1/200$ solar for the two 
components at $z \sim 1.92$ and $\approx 40$ \% solar 
at $z \sim 1.94$. 
%
\vskip 12pt 
Here we take advantage of new complementary data of high
resolution and high signal-to-noise ratio ($S/N$) to
further constrain metallicity inhomogeneities in these
systems. 
The paper is structured as follows: in Sect.~2 we report
observational details and the line fitting procedure
for the systems; in Sect.~3 we investigate the possible
ionising mechanisms and the determination of metallicity
and abundances. In Sect.~4, the observed behaviour of the 
ratio \siq/\cq\ is described; finally, we report our
conclusions in Sect.~5.   


\section{Comments on individual systems}


\begin{figure}
   \begin{center}
    	\resizebox{\hsize}{13cm}{\includegraphics{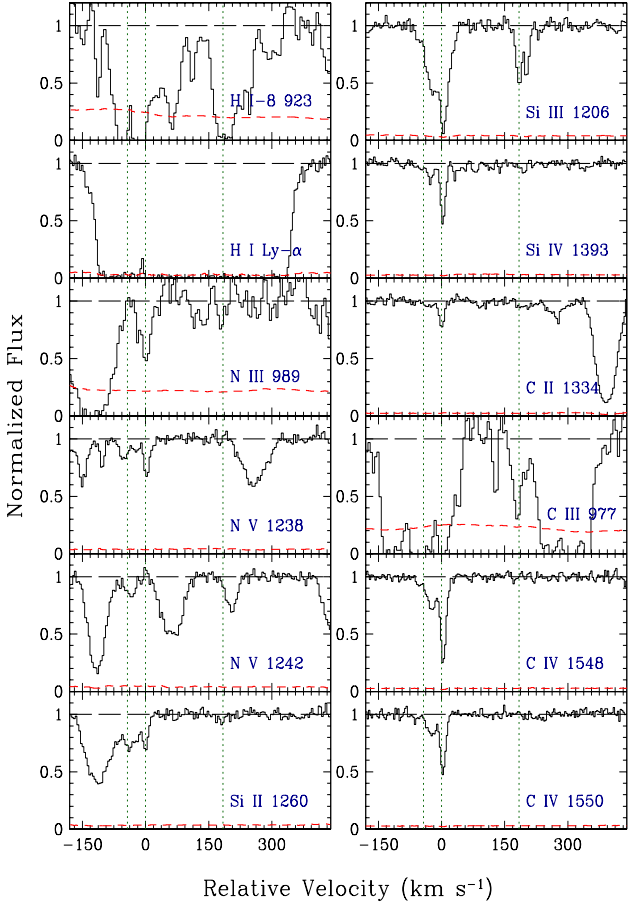}}
   \end{center}
   \caption{Main ionic transitions observed for the
    	system at $z\sim 1.92$. The dotted lines mark
    	the position of the clouds discussed in the text.
    	From left to right: cloud 1 at $v\simeq -43$
    	\kms\ ($z= 1.9255$) with metallicity
    	$\sim 1/500$ solar; cloud 2 at $v=0$ \kms\
    	($z=1.9259$) with metallicity $\sim 1/8$ solar
    	and cloud 3 at $v\simeq+184$ \kms\
    	($z=1.92577$) with metallicity $\sim 1/100$ solar} 
\label{fig:sys192}
\end{figure}

\begin{table}
\begin{center}
\caption{Ionic column densities for the $z\sim1.92$
system. An (f) indicates that in the fit the parameter
has been fixed to the value obtained for the previous
absorption line}  
\label{tab:sys192} 
\begin{tabular}{lcclc}
\hline\hline \\
      &      & $b$           &    & $\log N$ \\ 
Comp. & $z$  & (km s$^{-1}$) &Ion & (cm$^{-2}$) \\
&&&& \\
\hline \\
1\ldots & 1.92543$^a$ &  $11\pm 5$ & \siq  & $11.7\pm 0.2$ \\        
  &          &     & \sitre& $11.2\pm 0.2$ \\
  &          &     & \cq  & $12.2\pm 0.2$ \\   	 
2\ldots & 1.925719$^b$ &  $14\pm 1$  & \siq & $12.23\pm 0.05$ \\
  & 	      &        & \sitre &   $12.61\pm   0.04$ \\
  &           &           & \cq & $13.07\pm0.03$ \\
  & (f)       &  (f)      & \nc & $< 12.75\pm 0.08$ \\
  & (f)       &  (f)      & \ntre & $13.5\pm 0.2$ \\
  & 1.925718$^b$ & $18\pm11$ & \cdue &  $12.4\pm0.2$ \\
3\ldots & 1.925950$^c$  & $7.6\pm 0.6$ & \sidue & $12.26\pm0.03$ \\ 
  &          &         & \cdue & $12.91\pm0.05$ \\
  &          &         & \mgd & $11.72\pm 0.08$ \\
  & 1.925989$^c$  & $6.7\pm 0.2$ & \siq & $12.72\pm0.01$ \\
  &           &         & \sitre    & $12.76\pm 0.09$ \\
  &           &         & \cq       & $13.41\pm0.01$ \\
  &    (f)    &  (f)    & \nc       & $<12.4\pm0.1$ \\
  &    (f)    &   (f)   & \ntre     & $13.6\pm0.2$\\
4\ldots & 1.927747$^c$ & $7.4 \pm0.8$  & \sitre & $12.32\pm0.03$ \\
5\ldots & 1.927929$^c$ & $5.5 \pm0.8$  & \sitre & $12.14\pm0.04$ \\
\hline
\end{tabular}
\end{center}
$^a$ Error $30\times 10^{-6}$, $^b$ Error $7\times 10^{-6}$, $^c$
Error $4\times 10^{-6}$
\end{table}

\begin{figure}
   \begin{center}
    	\resizebox{\hsize}{13cm}{\includegraphics{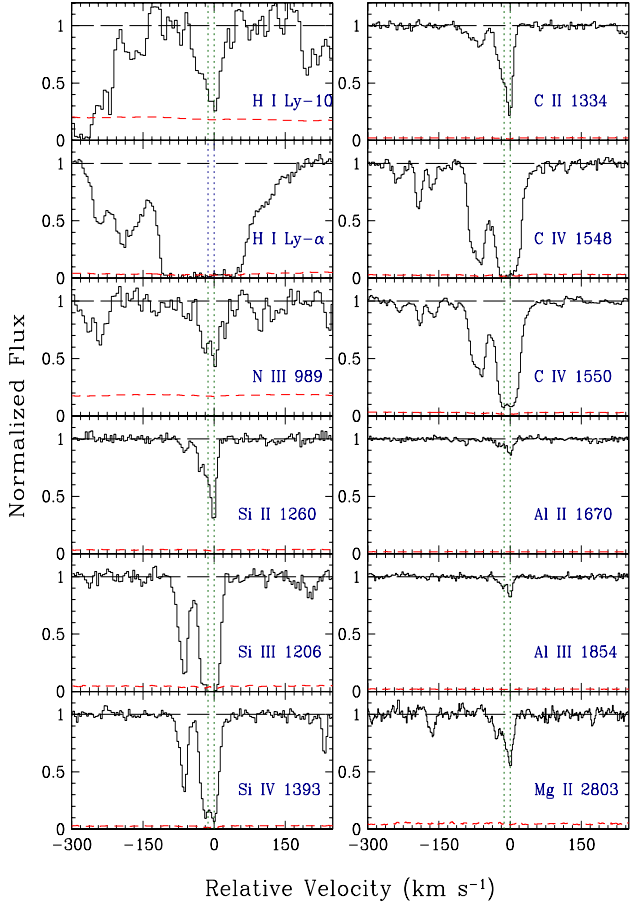}}
   \end{center}
   \caption{Main ionic transitions observed for the
    	system at $z\sim 1.94$. The origin of the
    	velocity axis is fixed at $z=1.942616$ and the
    	dotted lines indicate the positions of components
    	7 and 8 in Table~\ref{tab:sys194}} \label{fig:sys194}
\end{figure}

\begin{table}
\begin{center}
\caption{Ionic column densities for the $z\sim 1.94$ system
\label{tab:sys194}} 
\begin{tabular}{lcclc}
\hline\hline \\
      &     &  $b$          &     & $\log N$  \\ 
Comp. & $z$ & (km s$^{-1}$) & Ion & (cm$^{-2}$) \\
&&&& \\
\hline \\
1\ldots	& 1.94031$^a$  & $12  \pm 1  $	& \cq   &$12.67\pm0.04$ \\
	& 1.940258$^b$ & $19  \pm 1  $	& \osei &$13.9 \pm0.1$ \\
        &              &                & \huno &$13.28\pm0.03$ \\ 
2\ldots	& 1.940730$^b$ & $7.0 \pm 0.4$	& \cq   &$12.97\pm0.02$ \\
        &              & $22  \pm 3  $	& \osei &$13.9 \pm0.2$ \\
        &              &                & \huno &$13.53\pm0.07$ \\
3\ldots	& 1.941006$^b$ & $10  \pm 1  $	& \cq   &$12.78\pm0.03$ \\
        & 1.94107$^c$  & $22  \pm 4  $	& \osei &$14.1 \pm0.1$ \\
        &              &                & \huno &$13.2 \pm0.1$ \\  
4\ldots	& 1.941791$^b$ & $5.8 \pm 0.6$	& \cq   &$12.98\pm0.05$ \\
5\ldots	& 1.941975$^b$ & $14.3\pm 0.7$	& \siq  &$12.78\pm0.03$ \\
	&              &	        &\sitre &$12.70\pm0.03$ \\
	&     (f)      & $16.4\pm 0.6$	& \cq   &$13.89\pm0.01$ \\
	&  1.94182$^c$ & $20  \pm 4  $	& \cdue &$12.9  \pm0.1$ \\
	&     (f)      & $91\pm25$      & \huno &$13.80\pm0.07$ \\
6\ldots	& 1.942001$^b$  & $3.7 \pm 0.5$	& \siq  &$12.66\pm0.04$ \\
	&               &	        &\sitre &$12.4 \pm0.1$ \\
        &    (f)        & $5   \pm 1  $	& \cq   &$12.84\pm0.08$ \\ 
	& 1.942011$^b$  & $9   \pm 2  $	& \cdue &$12.8 \pm0.1$ \\
        &               &       	&\sidue &$11.79\pm0.08$ \\
	&    (f)        & $23.9\pm0.5$  & \huno &$15.60\pm0.06$\\
7\ldots	& 1.942429$^b$  & $10.3\pm 0.4$	& \siq  &$13.34\pm 0.02$ \\ 
	&               &	        &\sitre &$13.20\pm 0.04$ \\
	&     (f)       &     (f)       & \ntre &$13.7\pm 0.1$ \\
	&     (f)       & $10.1\pm 0.4$	& \cq   &$14.01\pm 0.03$ \\
	& 1.942484$^b$  & $17.1\pm 0.6$	& \cdue &$13.62\pm 0.02$ \\
	&               & 	        &\sidue &$12.59\pm 0.03$ \\
        &               &               & \mgd  &$12.54\pm 0.04$ \\
	&               &               &\altre &$11.92\pm 0.05$ \\
 	&               &               &\aldue &$11.39\pm 0.05$ \\
	&    (f)        & $26\pm 2 $    &\huno  &$16.1 \pm 0.1$ \\
8\ldots	& 1.942618$^b$  & $9.1 \pm 0.03$& \siq  &$13.39\pm 0.02$ \\
	&               &               &\sitre &$13.46\pm 0.05$ \\
	&     (f)       &     (f)       & \ntre &$13.8\pm 0.1$ \\
	&     (f)       & $14  \pm 1  $ & \cq   &$14.1 \pm0.1$ \\ 
	& 1.942614$^b$  & $5.2 \pm 0.3$ & \cdue &$13.52\pm0.03$ \\
	&               &               &\sidue &$12.53\pm 0.03$ \\
	&               &               & \mgd  &$12.37\pm0.05$ \\
        &               &               &\altre &$11.79\pm0.05$ \\
	&               &               &\aldue &$11.21\pm0.06$ \\
        &     (f)       & $15\pm 4$     &\huno  &$15.8\pm 0.2$ \\
9\ldots	& 1.94269$^c$   & $18  \pm 1$   & \cq   &$13.8 \pm0.2 $\\
\hline 
\end{tabular}
\end{center}
$^a$ Error $11-20 \times 10^{-6}$, $^b$ Error  $\le 10 \times
10^{-6}$, $^c$ Error  $> 30 \times 10^{-6}$ 
\end{table}

\begin{figure}
   \begin{center}
	\resizebox{\hsize}{13cm}{\includegraphics{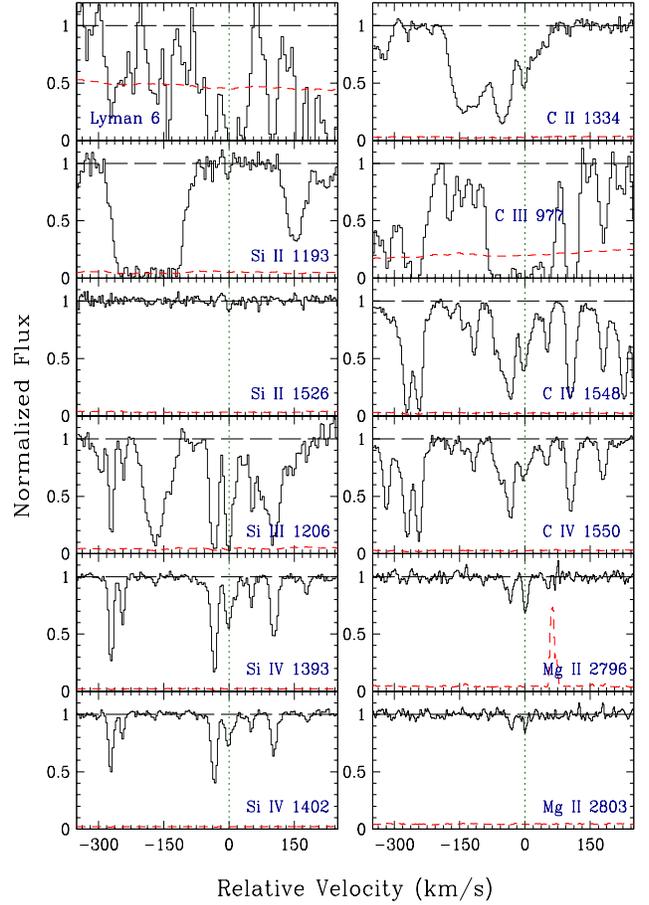}}
   \end{center}
   \caption{Main ionic transitions for the system at $z
	\sim 1.87$. The origin of the velocity axis is
	fixed at $z=1.87008$ corresponding to component
	12 analysed in the text} \label{fig:sys187}	
\end{figure}


New data on the HDF-South quasar have been 
obtained in October 1999 during the commissioning 
of UVES, the UV and Visual Echelle Spectrograph,
mounted on the VLT Kueyen ESO telescope at Paranal
(Chile).  
The spectrum is of high resolution ($R=45000$) and
high signal-to-noise ratio ($S/N = 40-60$ per
resolution element) and covers the 
spectral range $\lambda\lambda\, 3050-10000$ \AA. 
The data reduction and spectrum analysis are reported 
in \citet{papI}. 
We use these data together with the echelle spectrum 
($\lambda\lambda\,2275-3118$ \AA) of high resolution 
($R=30000$) obtained with the STIS instrument on board
HST \citep{sav98}. 

When not otherwise stated, low-ionisation, weak lines of
different species have been fitted together with the same
number of sub-components, each with the same redshift and
Doppler width.
The same is true for high ionisation,
complex and/or saturated lines. 
Absorption lines detected in the STIS spectrum 
(mainly \ntre, \ctre) are fitted using the same
components used to fit \siq. 
%
Voigt profile fitting is obtained in the context Lyman of MIDAS
the ESO reduction package \citep{font:ball}. 


\subsection{The system at $z\sim 1.92$}


\begin{figure}
   \begin{center}
	\resizebox{\hsize}{6cm}{\includegraphics{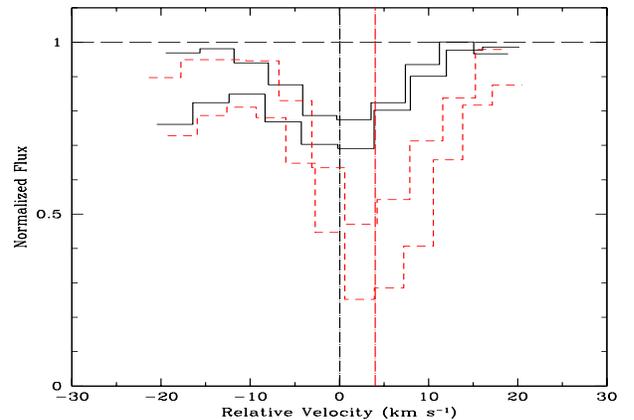}}
   \end{center}
   \caption{Superposition of the absorption lines due to \sidue\
	$\lambda\,1260$, \cdue\ $\lambda\,1334$ (solid
	line) at $z\simeq 1.92595$ and \siq\
	$\lambda\,1393$, \cq\ $\lambda\,1548$
	(short-dashed line) at $z \simeq 1.92599$. 
	The dot-dashed lines indicate
	the center of the two groups of lines separated by 4 \kms
	\label{fig:shift}} 	
\end{figure}

The strong, well defined component at $z\simeq 1.92599$
(see Fig.~\ref{fig:sys192} and Table~\ref{tab:sys192}) in
the high-ionisation line profiles (\cq, \sitre, \siq) is
displaced by $4 \pm 0.4$ \kms\ with respect to the 
corresponding component at $z\simeq 1.92595$ seen in 
the low ionisation lines (\cdue, \sidue) as shown in
Fig.~\ref{fig:shift}. 
The number of lines available together with the very
good quality of the data give confidence that this shift
is real. 
The simplicity of the velocity profiles suggests that
this reflects the internal structure of an \ion{H}{ii}
region flow in which kinematics and ionisation state are 
well correlated \citep[e.g.][]{HO99,rsb99}. 

The simultaneous fit of the \huno\ \Lya, \Lyb, 
Ly$\epsilon$ and Ly-8 lines results in two main
components at $z \simeq 1.92553$ and $z \simeq 1.92775$
with  $\log N$(\huno)$= 17.1 \pm 0.2$ and  $16.7\pm0.3$ 
respectively. 
Thus the two stronger \huno\ absorptions are shifted 
relative to the main \cdue\ and \sidue\ component 
observed at $z = 1.92595$ by $\sim -43$ \kms\ and $+184$
\kms\ respectively.  
The mismatch in the velocity positions of the strongest
\huno\  and \cdue\ absorptions is apparent in
Fig.~\ref{fig:sys192}, where the origin of the velocity
axis is fixed at the position of the \cdue\ absorption.
The upper limit to the \huno\ column density at the
redshift of the main metal component is 
$\log N$(\huno)$\lsim 16$. This is derived by adopting
the Doppler parameter as measured from
the \cdue\ line. 
It can be seen as well that at velocity position 
+184 \kms, the only metal transition detected is 
\sitre.  
The  $3\, \sigma$ upper limit on the corresponding 
\cq\ absorption is $\log N$(\cq)$\lsim 11.9$. 


 


\subsection{The system at $z\sim 1.94$}


\begin{figure}
   \begin{center}
	\resizebox{\hsize}{!}{\includegraphics{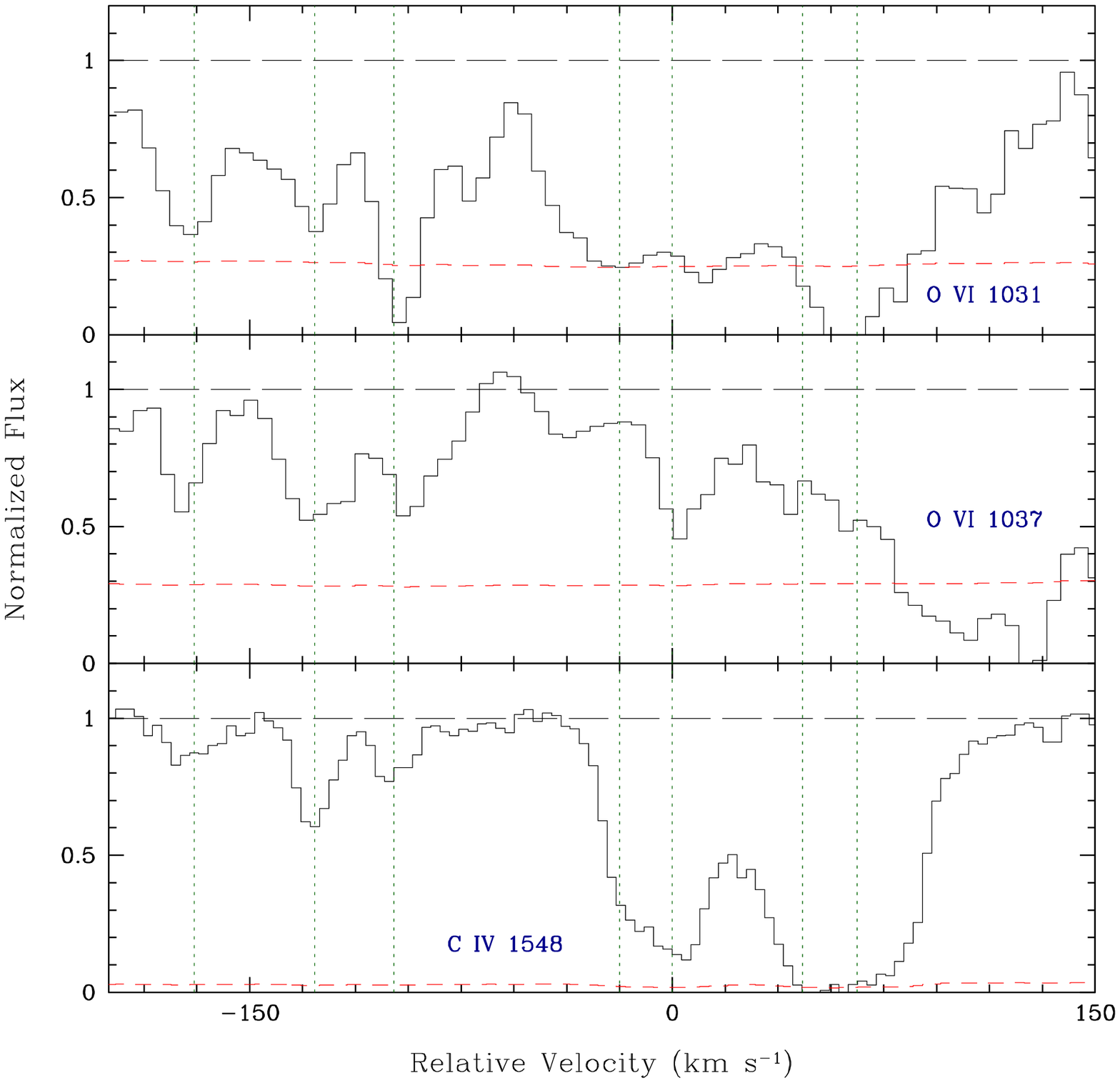}}
   \end{center}
   \caption{ \osei\ $\lambda\lambda\,1031,1037$ for the system at
	$z\sim1.94$. The velocity axis is centered at
	$z=1.941975$. The dotted lines show the position
	of the main components in \cq \label{fig:osei194}}	
\end{figure}

Absorption profiles for this system are shown in
Fig.~\ref{fig:sys194}. 

\siq\ and \sitre\ absorptions are fitted with the
same components and Doppler parameters. 
\cq\ is saturated; we therefore fix the components at
the redshifts determined for \siq\ and then add weak 
components to reach a good fit. 
\sidue\ $\lambda\,1193,\ \lambda\,1260,\ \lambda\,1304,\
\lambda\,1526$,  \cdue\ $\lambda\,1334$, \mgd\
$\lambda\,2803$,  \aldue\ $\lambda\,1670$ and \altre\
$\lambda\,1854$ are fitted together (same redshift
and same Doppler parameter). \mgd\ $\lambda\,2796$ is 
blended with a telluric absorption and is not
considered in the fit. 
Results are given in Table~\ref{tab:sys194}.   
Most components shows significant differences in velocity 
(up to $\sim 25$ \kms) between high-ionisation and
low-ionisation transitions, but blending is likely the
main cause of this discrepancies. 

Associated \osei\ is detected in the STIS portion of the spectrum
(see Fig.~\ref{fig:osei194}). 
Although the $S/N$ ratio is not excellent, the \osei\ and
\cq\ profiles are correlated with in particular three
components (1, 2, 3 in Table~\ref{tab:sys194}) well
detected in both species at 
$z \simeq$ 1.94026, 1.94073, 1.94107 ($v = -175$, $-127$  
and $-92$ \kms  respectively in
Fig.~\ref{fig:osei194}).  

The \huno\ column densities quoted in Table~\ref{tab:sys194}
are obtained by fitting together \huno\ \Lya, Ly$\gamma$, 
Ly$\epsilon$, Ly-10, Ly-12 at the redshifts fixed by the
\cdue\ components.
\nc\ is not observed; the $3\,\sigma$ upper limit on the column
density is $\log N$(\nc)$\lsim 12.15$. 


\subsection{The system at $z\sim 1.87$}


This system has a complex structure spanning around 600
\kms, with 11 (16) components in \siq\ (\cq). \mgd\ is
observed only at the redshift of the two central
components (11 and 12 in Table~\ref{tab:sys187}).   
Unfortunately, important transitions like \cdue\
$\lambda\,1334$, \sidue\ $\lambda\,1260$, \sitre\ 
$\lambda\,1206$ and \ctre\ $\lambda\,977$ are
badly affected by blending (see Fig.~\ref{fig:sys187}).  

The \siq\ doublet is fitted by itself and the redshifts of 
the components are used to fit \sitre. 
\cq\ is fitted by itself because more components with
respect to \siq\ are necessary to reach an acceptable fit.
Table~\ref{tab:sys187} displays the fitting parameters for the 17
components. 

\begin{figure}
   \begin{center}
	\resizebox{\hsize}{!}{\includegraphics{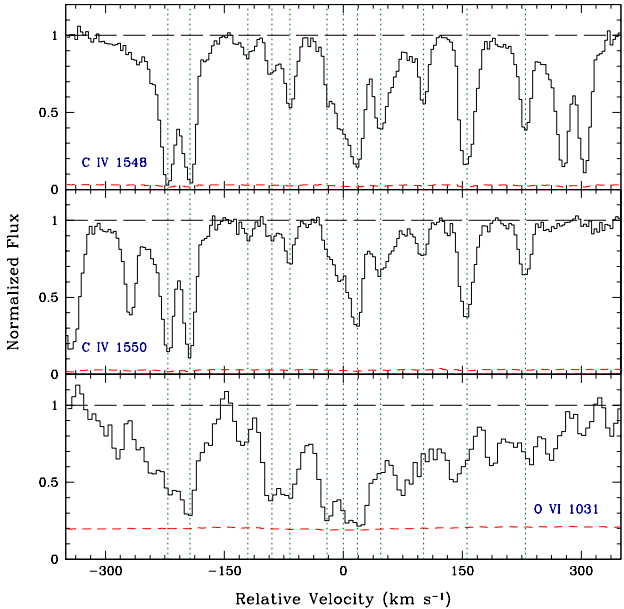}}
   \end{center}
   \caption{ \osei\ $\lambda\lambda\,1031,1037$ for the system at
	$z \sim 1.87$. The velocity axis is centered at
	$z=1.8695$. The dotted lines show the position of the main
	components in \cq \label{fig:osei187}}	
\end{figure}

Associated \osei\ is detected in the STIS portion of 
the spectrum (see Fig.~\ref{fig:osei187}), low $S/N$
ratio and partial blending with
intervening \huno\ lines prevent to carry on a
best fit procedure in MIDAS. 
An indicative  upper limit to the column densities 
in each component is estimated to be $\sim 13.7$     
using the interactive fitting program Xvoigt
\citep{xvoigt}. 
It is apparent from Fig.~\ref{fig:osei187} that, 
although the \cq\ and \osei\ profiles are correlated, 
the distribution of the \osei\ phase is smoother than 
that of the \cq\ phase.
\nc\ is not detected: the $3\,\sigma$ limit on the column 
density is: $\log N$(\nc)$<12.3$ for $z < 1.87$ and 
$\log N$(\nc)$<12.6$ for $z > 1.87$.  
 
Due to the heavy saturation of \Lya\ and \Lyb\
absorptions and to the complexity of the system, it is
not  possible to reliably constrain \huno\ column
densities for all the components.
For components 2, 4 and 16 we estimate $\log N$(\huno)$\simeq
15.2\pm0.2$, $\simeq 14.9\pm0.1$ and $\simeq 15.4\pm0.1$
respectively from the simultaneous fit of \huno\ \Lya,
\Lyb, Ly$\gamma$, and Ly-6.


\section{Ionisation models}


The nature of the ionisation mechanisms is investigated
comparing the observed column densities with 
those predicted by photoionisation models. 
For this we use the Cloudy software package 
\citep{cloudy} assuming ionisation equilibrium.
In the models we adopt: (1) a total hydrogen density
$n$(H)$= 10^{-2.5}$ cm$^{-3}$, (2) a plane parallel
geometry for the gas cloud, (3) the solar abundance
pattern, and two ionising spectra (4a) the Haardt-Madau 
\citep[][hereafter HM]{HM96,MHR99} EUVB spectrum with galaxy   
contribution or (4b) a composite (hereafter SSP) of  
the spectrum produced by a simple stellar population with 
20 \% solar metallicity, age 0.1 Gyr 
(Charlot private communication) plus a power law, $f(\nu)
\propto \nu^{-1}$, which increases the flux at energies
greater than the \ion{He}{ii} ionisation break (54.4 eV)
by a factor of $\sim 3$ \citep{sk97}.    

The ionisation parameter, $U$, defined by 
\begin{equation} 
U \equiv \frac{\phi_{912}}{cn_{\rm H}}=\frac{4\pi J_{912}}{hcn_{\rm
H}}= 2 \times 10^{-5} \frac{J_{912}/10^{-21.5}}{n_{\rm H}/(1
\mbox{cm}^{-3})}
\end{equation} 
\noindent
(where $J_{912}$ is the intensity of the EUVB at 1 ryd) and the
metallicity of the system are derived by matching the model output
column densities with the observed values. 
 

\subsection{The system at $z \sim 1.92$}


In Sect.~2.1 we have singled out three clouds at  
$z\simeq$ 1.9255, 1.9259, 1.9277 with $\log N$(\huno)
$\simeq 17.1$,  $<$16, 16.7 respectively, corresponding 
to the $v = -43$ \kms\ component which is the main 
component in \huno\ (cloud 1), the $v =0$ \kms\ component 
which is the main component in \cdue\ and \sidue\ (cloud
2)  and the $+184$ \kms\ component which is detected in 
\huno\ and \sitre\ only (cloud 3). 
At variance with PB99, we anlyse separately cloud 1
and cloud 2. As there is an uncertainty of $\sim 20$
\kms\ on the redshift determination of cloud 1, we use
for it the metal column densities of the slightly shifted
component 2 at $z\simeq 1.92572$ (see
Tab.~\ref{tab:sys192}). Thus, the metallicity we
determine for this cloud could be an upper limit. 
Finally, the model for cloud 3 takes into account the
possible \sitre\ absorption line (component 4 in
Tab.~\ref{tab:sys192}) which was not observed by PB99.

Because of the large number of parameters
(ionisation parameter, shape of the ionising flux, 
metallicities), the metallicity determination is most
often degenerate.  
However, when the number of transitions is large and the
column densities  well determined, the mean metallicity
and ionisation parameter can be estimated with reasonable 
uncertainties. 

\begin{figure}
   \begin{center}
	\resizebox{\hsize}{!}{\includegraphics{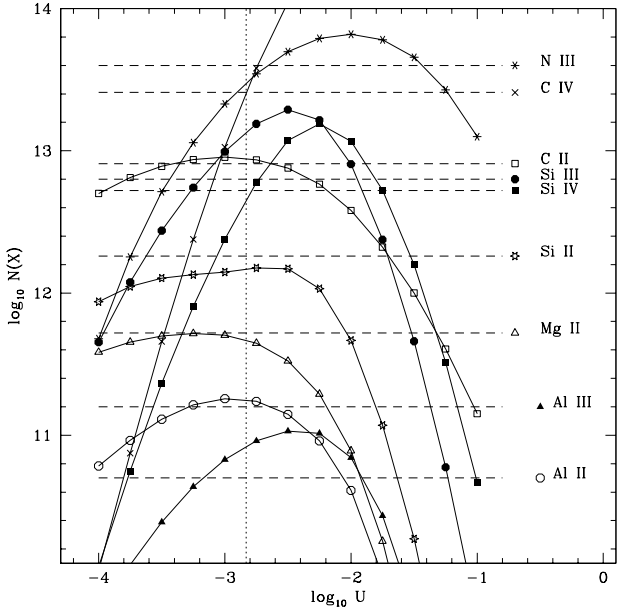}}
   \end{center}
   \caption{Ionic column densities computed by Cloudy (Ferland, 1997) 
   for  $\log N_{\rm HI} = 16$,  [X/H]~$= -0.87$, [Si/H]~$= 0.15$
   and an SSP ionising spectrum with 20 \% solar metallicity (see
   text) vs. a modified ionisation parameter, $U$, as defined in the
   text. The symbols on the right side of the plot and the dashed
   orizontal lines indicate the observed column density
   values for cloud 2. 
   The symbols sifhted to the right are upper limit measurements. 
   The vertical dotted line indicates the best value of $U$ determined
   from the observed \cq, \cdue, \siq, \sitre and \sidue\
   column densities \label{fig:SSP192}} 
\end{figure}

The observed column densities are reproduced by 
a SSP spectrum and a slight overabundance of silicon  
([Si/H]~$\sim 0.15$). 
The higher quality data allow us to reject the HM
spectrum model which was considered viable in the
analysis by PB99.
Indeed, the observed  column density pattern 
(\cdue, \cq, \sidue, \sitre, \siq) of cloud 2 is not
reproduced by the HM spectrum for any combination of 
metallicity and ionisation parameter, even  
allowing for an over-solar abundance of silicon.
Fig.~\ref{fig:SSP192} shows the predicted column
densities as a function of $U$ for cloud 2 models. 
The column densities obtained in our best 
model are given in Table~\ref{tab:SSP192}. The computed
values for silicon ions can still be reconciled with the
observed profiles. The discrepancy between observed
and predicted \nc\ and \ntre\ column densities, discussed
in PB99, is no longer present in our analysis.    
It is apparent from Fig.~\ref{fig:SSP192}  that the
\aldue\ column density is predicted larger than the upper
limit derived from observations. This behaviour is even
more apparent in Fig.~\ref{fig:HM194} both for the column
densities of \altre\ and \aldue\ and for their ratio
which is predicted $\sim 0.3$ dex lower than observed.  
The underabundance of aluminium observed in galactic
low metallicity stars could reconcile the observed
and computed results.  
On the other hand, magnesium does not seem to be
overabundant as would be expected from the same halo star
observations.  
Furthermore, the recombination coefficients used to
compute the aluminium ionisation equilibrium are probably
questionable as already noted by \citet{ppj94}.    

For the other clouds, we derive that cloud 1 has similar 
ionisation parameter as cloud 2, $\log U \simeq -2.7$,
$-2.8$. The metallicities of the two clouds
differ by two  order of magnitudes: [X/H] $\simeq -2.7$ and
$-0.9$  respectively. 
The computation for cloud 3 is based solely on the
\sitre\ column density and on the upper limits on the
main transitions of silicon and carbon. 
The gas is probably in a low ionisation state, 
$-3.3 \lsim \log U \lsim -3.1$, and has a low metallicity
$-2.3 \lsim$ [X/H] $\lsim -2.0$. 
%
%
%
%
%
\begin{table}
\begin{center}
\caption{Observed and predicted ionic column densities
for cloud 1 in the system at $z\sim 1.92$ (see text) and
component 8 in the system at $z\sim 1.94$
\label{tab:SSP192} } 
\begin{tabular}{lcccc}
\hline\hline \\
& \multicolumn{2}{c}{cloud 1 ($z\simeq1.92595$)} &
\multicolumn{2}{c}{comp. 8 ($z\simeq1.942618$)} \\
Ion & $\log N_{\rm obs}$ & $\log N_{\rm pred}$ & $\log
N_{\rm obs}$ & $\log N_{\rm pred}$ \\
&&&& \\
\hline \\
\cdue  & $12.91\pm 0.05$  & 12.94 &  $13.50\pm 0.03$ & 13.5 \\               
\cq    & $13.41\pm0.01$   & 13.41 &  $14.1 \pm 0.1$  & 13.8 \\        
\ntre  & $13.6 \pm 0.2$   & 13.48 &  $13.8 \pm 0.1$  & 14.2  \\   
\nc    & $< 12.4\pm 0.1$  & 11.55 &  $<  12.15 $     & 11.56  \\       
\mgd   & $11.72 \pm 0.08$ & 11.67 &  $12.37\pm 0.05$ & 12.39  \\       
\altre & $< 11.19$ &        10.92 &  $11.79\pm 0.05$ & 12.09  \\      
\aldue & $< 10.66$ &        11.25 &  $11.21\pm 0.06$ & 11.82  \\      
\sidue & $12.26\pm 0.03$  & 12.16 &  $12.59\pm 0.04$ & 12.59 \\             
\sitre & $12.8 \pm 0.1$   & 13.1 &  $13.46\pm 0.05$ & 13.81  \\      
\siq   & $12.72\pm 0.01$  & 12.64 &  $13.39\pm 0.02$ & 13.39 \\             
\hline
\end{tabular}
\end{center}
\end{table}

%
\vskip 12pt

From this we derive that most probably the three clouds
are ionised by a local source of ionisation.
This conclusion is strengthened by the observed
correlation between kinematics and ionisation state in
the component at $z\sim 1.9259$ (cloud 2). 
This together with the observations of variations in the
metal content by about two orders of magnitude over 
a velocity interval of $\sim 50$ \kms, suggests
that the line of sight crosses a region of intense
star-formation activity. 
It would be most interesting to probe this by 
searching the field for emission line objects at this
redshift. 


\subsection{The system at $z \sim 1.94$}


Photoionisation models have been constructed for the two
strong components 7 and 8 (see Table~\ref{tab:sys194}) 
both with $\log N$(\huno) $\approx 16$.
We have used as constraints the column densities of
\sidue, \cdue, and \siq. We do not use
\cq\ and \sitre\ because the lines are saturated.  

\begin{figure}
   \begin{center}
    	\resizebox{\hsize}{!}{\includegraphics{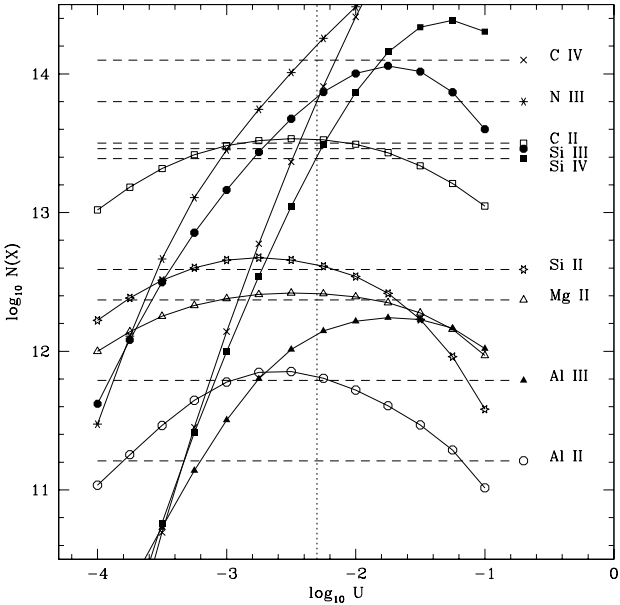}}
   \end{center}
   \caption{Results from a series of Cloudy (Ferland,
   1997) calculations assuming an overabundance 
   [Si/H]~$\simeq0.2$, a metallicity [X/H]=-0.5, $\log
   N_{\rm HI} = 15.8$ and an HM ionising spectrum
   including the 
   contribution of galaxies. Plotted are the calculated ionic column
   densities vs. a modified ionisation parameter, $U$, as defined in
   the text. The symbols on the	right side of the plot and the dashed
   orizontal lines indicate the observed column density values for the
   component of the metal system at
   $z=1.942614$, symbols shifted to the right give upper limit
   measurements. The vertical dotted line indicates the best value of
   $U$ determined from the observed \cdue, \siq\ and \sidue\ column
   densities \label{fig:HM194}} 
\end{figure}

We cannot find any satisfactory fit to the data for a SSP
ionising spectrum, except allowing for a Si/C ratio
smaller than solar which is probably unreasonable.
On the other hand, the HM spectrum accounts for the
observed column densities in both components for $\log
U\sim -2.3$ and  $-0.7 \lsim $~[X/H]~$\lsim -0.5$. 
The uncertainty on the metallicity is mostly due to the
uncertainty on the \huno\ column density determination.  
The model requires also an overabundance of silicon,
[Si/H]~$\simeq0.2$.  
In Fig.~\ref{fig:HM194} the computed column densities for all 
the observed ions as a function of $U$ are reported in
the case of component 8 and the values for the best fit
parameters are in Table~\ref{tab:SSP192}. It can be seen that most of the 
column densities are explained within a factor of two which is 
probably acceptable given the uncertainties on the column densities.


As previously noted, there is a discrepancy between the 
observed and computed \aldue\ and \altre\ column
densities. 
%
%
\ntre\ is slightly overpredicted. Nitrogen could 
be underabundant which would not be surprising in the 
hypothesis of secondary origin of this element. 
A firm conclusion cannot be reached as the \ntre\ column
density is not well determined.

\vskip 12pt

The three weak components 1, 2, and 3 show absorption   
due to \osei, \cq\ and \huno.
The observed ratios log~(\cq/\osei), log~(\cq/\huno)  
are similar for the three components, average values are
$\simeq -1.1 \pm 0.2$ and $-0.5\pm0.1$
respectively.  
We investigate both the possibilities of    
photo- or  collisional ionisation. 

The two adopted ionising spectra can reproduce the
observed column density pattern in  
the context of photoionisation.
But \nc\  is always overpredicted: 
$\log N$(\nc)$ \sim 13$ instead of $<$12.15 (3$\sigma$)
as observed.   
In the case of the HM spectrum the parameters vary in the
ranges $-0.15 \lsim \log U \lsim 0.2$ and $-1.4
\lsim$~[X/H]~$ \lsim -1.1$  for the three components.
While for the SSP spectrum $-0.86 \lsim \log U \lsim
-0.55$ and $-0.52 \lsim$~[X/H]~$\lsim -0.22$.

In the hypothesis of collisional ionisation, the observed
column density ratio log~(\cq/\osei) 
is consistent with a temperature $\log T = 5.35$
\citep{su:do93}. 
The prediction for the other ratios are: 
log~(\cq/\huno)$ \simeq -0.03$, log~(\osei/\huno)$
\simeq 1.30$ for solar metallicities.   
In order to match the observed ratios, the metallicity
should be decreased to $\sim 1/3$ solar.
Thus, if we assume collisional ionisation for these
components and a solar abundance pattern, we get a
metallicity comparable to the one predicted by Cloudy
with the SSP spectrum.  
At this temperature, collisional ionisation predicts
also: log~(\nc/\cq)$ \sim 0.6$ while 
$-0.5$ is observed. Therefore in the framework of these
models, \nc\ observations can be explained only if
nitrogen is deficient by an order of magnitude compared
to carbon. 

\vskip 12pt

In conclusion, the gas in this system is likely of quite
high metallicity (larger than 0.1 solar) for this
redshift. The ionisation state can be explained by
photoionisation  by a HM type spectrum. 
In this case however, metallicity is three times less 
in the high-ionisation phase (components 1, 2 and 3) 
compared to the low-ionisation phase (components 7 and 8).
The model can be reconciled with similar metallicities in  
both phases in case the high ionisation phase is
predominantly  ionised by a local source of photons or if
it is collisionally ionised.

Both the photo- and collisional ionisation models predict
that nitrogen is underabundant with respect to solar
[N/C]$ \lsim -0.5$.  



\subsection{The system at $z\sim 1.87$}


We first determine an upper limit on the temperature of
the gas from the measured Doppler parameter, in the
hypothesis of pure thermal broadening. 
We choose three narrow, resolved components (2, 4 and 17;
see Table~\ref{tab:sys187} and Fig.~\ref{fig:sys187})
with detected \siq, \sitre, \cq\ and, the last one,
\ctre\ absorptions; the observed Doppler parameter are
consistent with a temperature $\log T < 4.7$. 
In the hypothesis of collisional ionisation, at this
temperature most of the carbon would be in \cdue\ and
\ctre\ ions, \osei\ would not be present, and silicon
would be mainly in \sitre\ and \siq. 
From the ionic fractions calculated by \citet{su:do93}, 
at $\log T = 4.6$ $\log$ (\sitre/\siq) $
\simeq 0.8$, and at  $\log T = 4.8$ $\log$ (\sitre/\siq)
$\simeq 0.16$ and $\log$ (\ctre/\cq)$ \simeq 1.74 $.   
This excludes the possibility that the gas phase
giving rise to the observed silicon and carbon
transitions is collisionally ionised. 

However, \osei\ could still arise from a  
collisionally ionised phase. Indeed, the co-existence of 
these two ionising mechanisms has been explained by
e.g. the presence of a tail of shock-heated gas at high 
temperature away from the curve of equilibrium
\citep{HSR96} or a two phase halo where clouds 
photoionised by the EUVB  at an equilibrium temperature of 
$\sim 10^4$ K and giving origin to the relatively narrow lines of 
\cq, \siq, etc., are in pressure equilibrium with 
the hotter halo gas in which \osei\ arises
\citep{ppj92,mo:mir96}.   
The latter scenario predicts broad, shallow absorption
for \osei\ and \cq\ originating in the hot gas. 
The low $S/N$ ratio STIS spectrum marginally
shows that these lines are not as broad as those observed
e.g. by \citet{kt99}. 



\begin{figure}
   \begin{center}
	  \resizebox{\hsize}{!}{\includegraphics{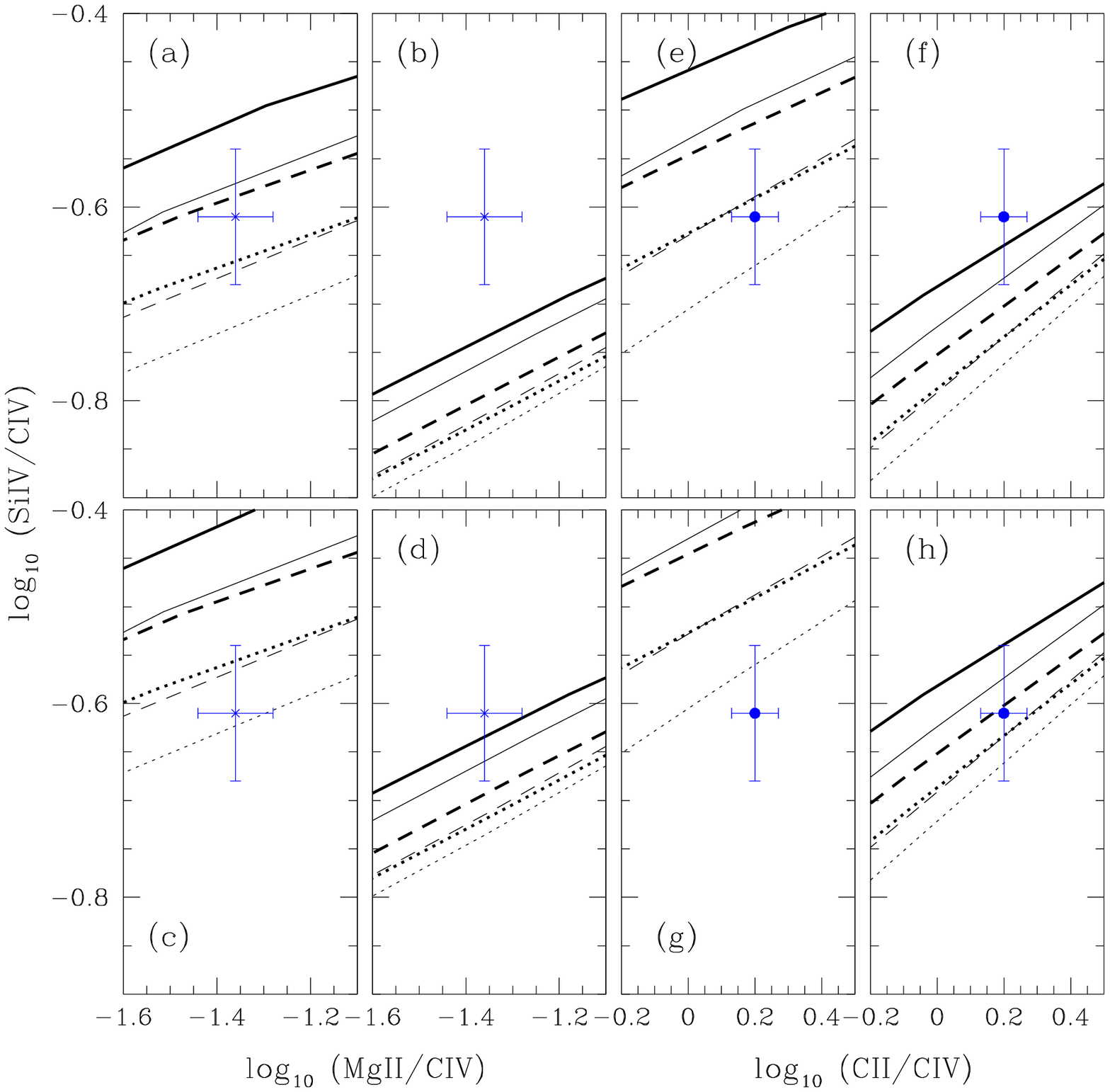}}
   \end{center}
   \caption{Results from a series of Cloudy calculations (Ferland, 
	  1997). Curves represents \siq/\cq\ column density ratio
	  vs. \mgd/\cq\ (panels {\bf a}-{\bf d}) and
	  \cdue/\cq\ (panels {\bf e}-{\bf h}) ratio as the
	  ionisation parameter varies in the range 
	  $-4 < \log U < -1$. Different curves are for
	  $\log N$(\huno)$ = 15$ (thin lines) and $\log N$(\huno)$ =
	  17$ (thick lines), [X/H]~$=-0.5$ (solid lines), [X/H]~$=-1$
	  (dashed lines), and [X/H]~$=-1.5$ (dotted lines). The 
	  panels have different ionising spectra and silicon
	  abudances: {\bf a},{\bf e} HM ionising spectrum
	  and solar abundance pattern; {\bf b},{\bf f}
	  SSP spectrum and solar abundance pattern; 
	  {\bf c},{\bf g} HM ionising spectrum and
	  [Si/H]~$=0.1$; {\bf d},{\bf h} SSP
	  spectrum and [Si/H]~$=0.1$ The superposed points
	  are the observed \mgd/\cq\ and \cdue/\cq\
	  column density ratios 
	  for component 12 of the system at $z\sim 1.87$} 
   \label{fig:grid187} 
\end{figure}

\begin{figure}
   \begin{center}
	  \resizebox{\hsize}{!}{\includegraphics{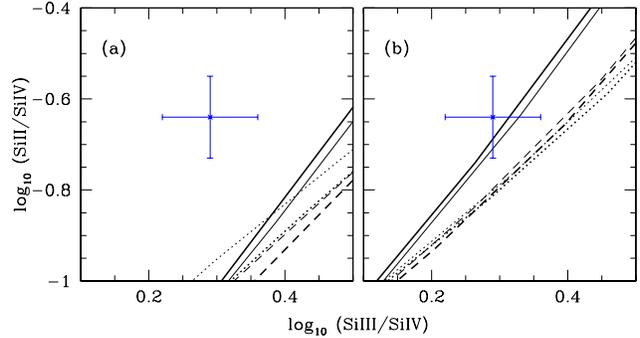}}
   \end{center}
   \caption{Results from a series of Cloudy calculations 
          (Ferland, 1997). Curves represent \sidue/\siq\
	  vs. \sitre/\siq\ column density ratio as the
	  ionisation parameter varies in the range $-4 <
	  \log U < -1$. The different curves describe 
	  the same models as in Fig.~\ref{fig:grid187}
	  for an HM ionising spectrum (panel {\bf a}) and
	  an SSP spectrum (panel {\bf b}).  
	  The superposed point is the corresponding
	  observed ratio}
   \label{fig:gridsi187} 
\end{figure}

The high uncertainty in the \huno\ column density
for most of the components and the blending of
many interesting absorption lines (refer to Sect.~2.3)
makes an abundance analysis for this system difficult. 
A fit to the column densities in component 2, at $z\simeq 
1.8675$, shows that the SSP spectrum model is best suited
with $\log U\simeq -2.36\pm0.04$, [X/H]~$\simeq
-0.4\pm0.1$ and an overabundance of silicon,
[Si/H]~$\simeq0.1$.  

In the case of component 12 at $z\simeq1.87008$ for which
we do not have a reliable determination of \huno\ column
density, we run a grid of photoionisation models 
varying the main parameters.
On the basis of what is found from the previous
modelisation, we consider metallicities between $\sim
1/3$ and $\sim 1/30$ solar, solar abundance pattern or 
an overabundance of silicon, [Si/H]~$=0.1$.  
Fig.~\ref{fig:grid187} shows the
predicted \siq/\cq\ ratio as a function of \mgd/\cq\ and
\cdue/\cq\ 
ratios for the ionising parameter varying 
between $-4 < \log U < -1$. 
Observed ratios are shown as black points. 
Three models are consistent with the data:
an HM ionising spectrum with solar abundance pattern and
metallicity [X/H]~$\sim -1$ for $\log N$(\huno)$=15$ or
[X/H]~$=-1.5$ for  $\log N$(\huno)$=17$ (panels {\bf
a},{\bf e});  an HM ionising spectrum with overabundance
of silicon,  $\log N$(\huno)$=15$ and metallicity
[X/H]~$=-1.5$ (panels {\bf c}, {\bf h}); 
an SSP spectrum with overabundance of silicon and a
metallicity [X/H]~$\sim -0.5$ (panels {\bf d}, {\bf g}).
The \sitre/\siq\ vs. \sidue/\siq\ plot favour the latter
model  (see Fig.~\ref{fig:gridsi187}).
 
\vskip 12pt

In conclusion, we find that the two components at $z\simeq 
1.8675$ and at $z\simeq 1.87008$ can be explained by a model 
with a local ionising stellar source and a gas
metallicity of $\sim 1/3$ solar. 
  
\section{The \siq/\cq\ ratio }

It is well known that the \siq/\cq\  ratio in
high-ionisation systems depends critically on the 
strength of the \ion{He}{ii} ionisation edge
(54 eV) break.  
Direct observations of the \ion{He}{ii} \Lya\ absorption
in QSO spectra \citep{jakob94,dkz96,har97,reim97,zdk98}
show a marked decrease of the opacity for $z < 3$. 
Detailed investigation of the \siq/\cq\ ratio have been 
carried out by \citet{song:cow96,sav97,gir:sh97} and
\citet{song98} among others. 
They all agree on the fact that the above ratio increases
strongly between $z=2$ and $z=4$ with a possible 
discontinuity around $z=3$. 
In particular, \citet{song98} collected a sample of 
\cq\ systems with $N$(\cq)$ > 5\times 10^{12}$ \cm, 
and obtained median values of $0.043^{+0.015}_{-0.008}$ for all the 
systems below $z=3$, and $0.15\pm0.04$ above $z=3$ (errors are
$1\,\sigma$ computed using the median sign method). 
\citet{boks98} questioned this result, based
on an analysis of the redshift evolution of the ion
ratios in individual Voigt components within
complex systems. 

\begin{figure}
   \begin{center}
	\resizebox{\hsize}{!}{\includegraphics{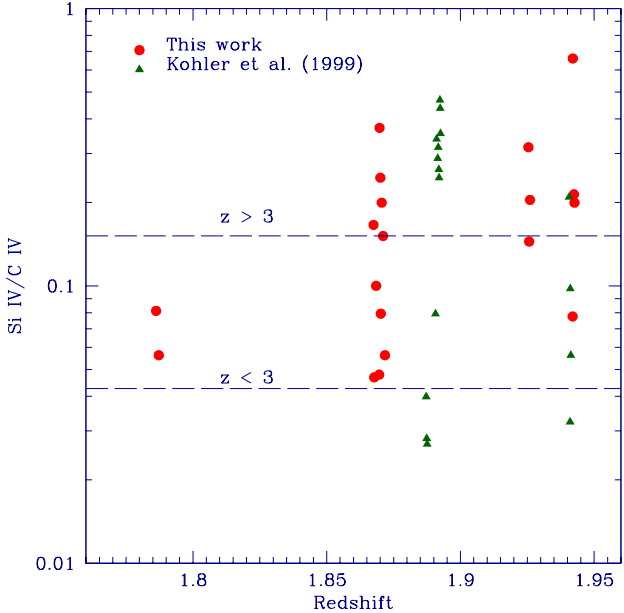}}
   \end{center}
   \caption{The values of the \siq/\cq\ column density ratio for the
	single components of the metal systems in our sample (see
	text) are reported as a function of redshift. The two
	long-dashed lines are the median values computed by Songaila
	(1998) from lines in her sample below and above $z=3$ } 
   \label{fig:sicar_z}	
\end{figure}

The \siq/\cq\ ratios for the individual components
of the three considered systems are plotted as a
function of redshift in Fig.~\ref{fig:sicar_z} together  
with the data by \citet{kohler99} who observed 4 
systems at similar resolution. All systems have $z<3$.
Data are spread over more than an order of magnitude and 
most of them are above the value estimated by
\citet{song98} for systems with $z<3$. 
It has to be noted that all but one of the systems in our
sample have total \cq\ column density $ > 10^{14}$ \cm, 
while only 10 \% of Songaila's systems at $z<3$ belongs
to this range of column densities.
 
As a comparison, we consider the HST data on absorption 
lines  arising in the Milky Way halo gas
\citep{savage00}. 
The observed \siq\ and \cq\
transitions are weak, thus we transform the observed
equivalent width in column density by using the curve of
growth relationship for optically thin gas.
The weighted mean of the 9 obtained estimates 
(corresponding to different lines of sight) is
\siq/\cq~$\simeq 0.413\pm0.009$ and the values are spread
between $\sim 0.1$ and $\sim 0.6$.

This result supports the idea that high column density
systems, which most probably arise in halos of galaxies, 
have higher values of the \siq/\cq\ ratio compared to 
weak systems arising in regions more typical of the
intergalactic medium. 
The systems studied here are probably 
influenced by local ionising sources rather than by 
the diffuse UV background. 

\section{Conclusion}

We have discussed the nature of the three prominent metal
absorption systems observed in the UVES spectrum of the
HDF-South quasar J2233-606. 

We analyse the velocity structure of the systems and
their ionisation and abundance properties mainly by means 
of photoionisation models performed with the Cloudy
software package \citep{cloudy}. 
The main conclusions we draw are the following: 

\vskip 12pt

\noindent
1. - The signature of a probable expanding flow in an
\ion{H}{ii} region is observed in the main metal
component of the Lyman limit system, at $z\simeq 1.9259$, 
which shows a correlation between ionisation state and
kinematics. Furthermore, the two main components in the
\huno\ absorption profile are shifted by $-43$ 
\kms\ and $+184$ \kms\ with respect to that component. 
The observed column density pattern is explained by a
photoionisation model with a local stellar source and
three different metallicities, $\sim 1/500$ and $\sim
1/100$ solar for the two \huno\ clouds and $\sim 1/8$
solar for the metal component. 
The high metallicity in the latter cloud and the
marked inhomogeneity suggest that the line of sight 
intersects a region of ongoing
star formation and recent metal enrichment.

\noindent
2. - The system at $z\sim 1.94$ shows high
metallicity, $\sim 1/4$ solar, when modelled
with photoionisation by an HM spectrum. There are three  
components at higher ionisation where strong \osei\
absorption is detected. The metallicity of these
components is lower or equal to the one in the low
ionisation components depending if the same or a
different ionising source (SSP spectrum or collisional)
is considered. 
\aldue\ and \altre\ are badly underestimated by the
model, this could be due to a still low accuracy in the
recombination coefficients for this element
\citep{ppj94}. 

\noindent
3. - The system at $z\sim 1.87$ shows associated \osei\
absorptions. The same ionisation origin for these lines
and the narrow transitions of \cq, \siq,\sitre, etc, is
excluded, thus a more complex scenario has to be taken
into account. In order to test the different explanations
which have been proposed it will be interesting to obtain
spectra at high resolution and signal-to-noise ratio of
the \osei\ complexes. 
We could investigate only two of the 17 components, both
agree with [X/H]~$\sim 1/3$ solar and a local stellar
ionising source. 

\noindent
4. - The analysis of the \siq/\cq\ column density ratio
for a sample of systems at $z\lsim2$ with total
$N$(\cq)$> 10^14$ \cm, shows a ditribution skewed towards
higher values than observed by \citet{song98}. A
comparison with analogous results for the gas in the halo
of our galaxy implies that higher
total \cq\ column density systems present higher values
of the ratio \siq/\cq, likely because they arise in a 
galactic environment and are influenced by local
ionising sources rather than by the UV background. 

\vskip 12pt

\begin{table}
\begin{center}
\caption{Ionic column densities for the $z\sim 1.87$ system
\label{tab:sys187}} 
\begin{tabular}{lcclc}
\hline\hline \\
      &     & $b$           &     & $\log N$ \\
Comp. & $z$ & (km s$^{-1}$) & Ion & (cm$^{-2}$) \\
&&&& \\
\hline \\
1\ldots & 1.86733$^c$ &$ 60\pm6   $& \cq    & $13.31\pm 0.09$	\\    
2\ldots & 1.867494$^a$  &$5.3\pm0.2 $& \siq   & $12.97\pm 0.01$	\\
        &    (f)      & $5.2\pm0.4$  & \sitre & $12.65\pm 0.03$	\\
	&    (f)      &    (f)       & \sidue & $11.7 \pm0.1$ \\
	&    (f)      &    (f)       & \cdue  & $12.4\pm0.1$ \\
        &             &$5.9\pm0.3 $& \cq    & $13.75\pm 0.02$	\\    
3\ldots & 1.86756$^b$  &$25 \pm2   $& \cq    & $13.63\pm 0.05$	\\    
4\ldots & 1.867748$^a$  &$4.6\pm0.4 $& \siq   & $12.49\pm 0.02$	\\
        &              &$5  \pm  1 $& \sitre & $11.93\pm 0.05$	\\
        & 1.867758$^a$  &$6. \pm0.3 $& \cq    & $13.82\pm 0.02$	\\    
5\ldots & 1.867980$^b$ &$1.5\pm0.7 $& \cq    & $12.  \pm 0.2 $	\\    
6\ldots & 1.868459$^b$ &$3  \pm3   $& \siq   & $11.6 \pm 0.1 $	\\
        &              &$ 9 \pm2   $& \cq    & $12.56\pm 0.05$	\\   
7\ldots & 1.868749$^b$ &$7.8\pm0.9 $& \cq    & $12.77\pm 0.04$	\\    
8\ldots & 1.868966$^a$ &$8.4\pm0.5 $& \cq    & $13.12\pm 0.02$	\\    
9\ldots & 1.869414$^a$ &$1.5\pm0.8 $& \cq    & $12.5 \pm 0.2 $	\\    
10\ldots& 1.86964$^c$  &$23 \pm4   $& \siq   & $12.3 \pm 0.1 $	\\
        &              &$18 \pm 1  $& \cq    & $13.62\pm 0.04$	\\   
11\ldots& 1.869767$^a$ &$6.6\pm 0.2$& \siq   & $13.12\pm 0.02$	\\
        &    (f)      &$8.4\pm 0.5$& \sitre & $13.09\pm 0.05$	\\
        & 1.869756$^b$ &$8  \pm1   $& \mgd   & $11.91\pm 0.04$	\\
        &             &        & \aldue     & $11.1 \pm 0.1 $	\\
	&             &        & \altre     & $< 11.48\pm0.09$ \\
	& 1.869779$^a$ &$7.7\pm0.5$ & \cq    & $13.55\pm 0.04$	\\    
12\ldots& 1.870075$^a$ &$8.3\pm0.6$ & \siq   & $12.72\pm 0.02$	\\
        &    (f)      &$6.5\pm0.5$ & \sitre & $13.01\pm 0.07$	\\
        & 1.870086$^a$ &$4.8\pm0.6$ & \sidue & $12.08\pm 0.09$	\\
        &             &           & \mgd    & $11.97\pm 0.03$	\\
	&             &           & \aldue  & $11   \pm 0.1 $	\\
	&             &           & \altre  & $<11.2\pm 0.1 $	\\
	&             &$10 \pm2  $ & \cdue  & $13.53\pm 0.04$	\\
        & 1.870061$^a$ &$ 12\pm1  $ & \cq    & $13.33\pm 0.07$	\\    
13\ldots& 1.870232$^b$ &$4  \pm1  $ & \siq   & $12.01\pm 0.08$	\\
        &    (f)      &$6  \pm1  $ & \sitre & $12.13\pm 0.06$	\\
        &             &$ 33\pm6  $ & \cq    & $13.1 \pm 0.1 $	\\    
14\ldots& 1.870411$^b$ &$3  \pm2  $ & \siq   & $11.6 \pm 0.1 $	\\
        &    (f)      &$5  \pm1  $ & \sitre & $11.7 \pm 0.1 $	\\ 
15\ldots& 1.870582$^a$ &$4.2\pm0.5$ & \siq   & $12.24\pm 0.03$	\\
        &    (f)       &           & \sitre  & $12.17\pm 0.07$	\\
        &              &$6.1\pm 0.7$ & \cq   & $12.94\pm 0.05$	\\   
16\ldots& 1.871079$^a$ &$8.6\pm 0.2$ & \siq  & $12.86\pm 0.01$	\\
        &    (f)      &           & \sitre  & $12.64\pm 0.07$	\\
        &    (f)      & $6.6\pm0.8$ & \ctre & $15.9 \pm 0.5 $	\\
        & 1.871102$^a$ & $10.2\pm 0.2$ & \cq & $13.68\pm 0.01$	\\
17\ldots& 1.871796$^b$ & $6\pm1 $   & \siq   & $12.03\pm 0.05$	\\
        &   (f)       &           & \sitre  & $11.6 \pm 0.1 $	\\
	&   (f)       & $11\pm3 $   & \ctre & $13.1 \pm 0.1 $	\\
        & 1.871808$^a$& $8.5\pm 0.3$  & \cq & $13.28\pm 0.01$	 \\   
\hline
\end{tabular}
\end{center}
$^a$ Error $\le 5 \times 10^{-6}$, $^b$ Error $6\div15 \times
10^{-6}$, $^c$ Error $\ge 50 \times 10^{-6}$  \\
\end{table}

\begin{acknowledgements}
We are pleased to thank St\'ephane Charlot and Marcella
Longhetti for providing us with the stellar spectrum
(SSP) we adopted in our photoionisation model. 
V.D. is grateful to Andrea Ferrara for useful
discussions. V.D. is supported by a Marie Curie
individual fellowship  from the European Commission under 
the programme ``Improving Human Research Potential and
the Socio-Economic Knowledge Base'' (Contract
no. HPMF-CT-1999-00029).  
\end{acknowledgements}

\bibliography{aamnem99,myref}

\begin{thebibliography}{}


\bibitem[\protect\astroncite{Bergeron \& Boiss\'e}{1991}]{berg91}
Bergeron, J., and Boiss\'e, P. 1991,
\newblock {A\&A}, {243}, 344

\bibitem[\protect\astroncite{Bergeron et al.}{1994}]{berg94}
Bergeron, J., Petitjean, P., Sargent, W.~L.~W., et al. 1994,
\newblock {ApJ}, {436}, 33

\bibitem[\protect\astroncite{Boksenberg}{1998}]{boks98}
Boksenberg, A. 1998, 
\newblock in P. Petitjean and S. Charlot (eds.), {Structure and
Evolution of the Intergalactic Medium from QSO Absorption Line
Systems}, Proc. of the 13th IAP Colloquium, pp. 85--90, Editions
Fronti\`eres 

\bibitem[\protect\astroncite{Chen \& Prochaska}{2000}]{cp00}
Chen, H.-W., and Prochaska, J.~X. 2000,
\newblock {ApJ}, {543}, L9

\bibitem[\protect\astroncite{Churchill et al.}{1996}]{church96}
Churchill, C.~W., Steidel, C.~C., Vogt, S.~S. 1996, 
\newblock {ApJ}, {471}, 164

\bibitem[\protect\astroncite{Churchill et al.}{2000a}]{church00a}
Churchill, C.~W., Mellon, R.~R., Charlton, J.~C., et al. 2000, 
\newblock {ApJS}, {130}, 91 

\bibitem[\protect\astroncite{Churchill et al.}{2000b}]{church00b}
Churchill, C.~W., Mellon, R.~R., Charlton, J.~C., et al. 2000, 
\newblock {ApJ}, {accepted}



\bibitem[\protect\astroncite{Cristiani \& D'Odorico}{2000}]{papI}
Cristiani, S., and D'Odorico, V. 2000, 
\newblock {AJ}, {120}, 1648

\bibitem[\protect\astroncite{Davidsen et al.}{1996}]{dkz96}
Davidsen, A.~F., Kriss, G.~A., Zheng, W. 1996, 
\newblock {Nat}, {380}, 47


\bibitem[\protect\astroncite{Ferland}{1997}]{cloudy}
Ferland, G.~J. 1997,
\newblock {Hazy, a brief introduction to Cloudy 94.00}, 
\newblock {http://www.pa.uky.edu/~gary/cloudy/}

\bibitem[\protect\astroncite{Fontana \& Ballester}{1995}]{font:ball}
Fontana, A., and Ballester, P. 1995, 
\newblock {ESO The Messenger}, {80}, 37

\bibitem[\protect\astroncite{Gardner et al.}{1997}]{gardner97}
Gardner, J.~P.,Katz, N., Hernquist, L., Weinberg, D.~H. 1997, 
\newblock {ApJ}, {484}, 31
 
\bibitem[\protect\astroncite{Giroux \& Shull}{1997}]{gir:sh97} 
Giroux, M.~L., and Shull, J.~M. 1997, 
\newblock {AJ}, {113}, 1505

\bibitem[\protect\astroncite{Guillemin \& Bergeron}{1997}]{guill}
Guillemin, P., and Bergeron, J. 1997, 
\newblock {A\&A}, {328}, 499

\bibitem[\protect\astroncite{Haardt \& Madau}{1996}]{HM96}
Haardt, F., and Madau, P. 1996,
\newblock {ApJ}, {461}, 20

\bibitem[\protect\astroncite{Haehnelt et al.}{1996}]{HSR96}
Haehnelt, M.~G., Steinmetz, M., Rauch, M. 1996,
\newblock {ApJ}, {465}, L95

\bibitem[\protect\astroncite{Henney \& O'Dell}{1999}]{HO99}
Henney, W.~J., and O'Dell, C.R. 1999,
\newblock {AJ}, {118}, 2350

\bibitem[\protect\astroncite{Hernquist et al.}{1996}]{hernquist96}
Hernquist, L., Katz, N., Weinberg, D., Miralda-Escud\'e, J.  1996,
\newblock {ApJ}, {457}, L51

\bibitem[\protect\astroncite{Hogan et al.}{1997}]{har97}
Hogan, C.~J., Anderson, S.~F., Rugers, M.~H. 1997, 
\newblock {ApJ}, {113}, 1505


\bibitem[\protect\astroncite{Jakobsen et al.}{1994}]{jakob94}
Jakobsen, P., Boksenberg, A., Deharveng, J.~M., et al. 1994,  
\newblock {Nat}, {370}, 35

\bibitem[\protect\astroncite{Katz et al.}{1996}]{katz96}
Katz, N., Weinberg, D.~H., Hernquist, L., Miralda-Escud\'e, J. 1996, 
\newblock {ApJ}, {457}, L57
 
\bibitem[\protect\astroncite{Kirkman \& Tytler}{1999}]{kt99}
Kirkman, D., and Tytler, D. 1999,
\newblock {ApJ}, {512}, L5

\bibitem[\protect\astroncite{K\"ohler et al.}{1999}]{kohler99}
K\"ohler, S., Reimers, D., Tytler, et al. 1999, 
\newblock {A\&A}, {342}, 395

\bibitem[\protect\astroncite{Madau et al.}{1999}]{MHR99}
Madau, P., Haardt, F., Rees, M.~J. 1999,
\newblock {ApJ}, {514}, 648

\bibitem[\protect\astroncite{Mar \& Bailey}{1995}]{xvoigt}
Mar, D.~P., and Bailey, G. 1995, 
\newblock {Proc. Astron. Soc. Aust.}, {12},  239

\bibitem[\protect\astroncite{Miralda-Escud\'e et al.}{1996}]{miralda96} 
Miralda-Escud\'e, J., Cen, R., Ostriker, J.~P., Rauch, M. 1996, 
\newblock {ApJ}, {471}, 582

\bibitem[\protect\astroncite{Mo \& Miralda-Escud\'e}{1996}]{mo:mir96}  
Mo, H.~J., and Miralda-Escud\'e, J. 1996, 
\newblock {ApJ}, {469}, 589 


\bibitem[\protect\astroncite{Petitjean et al.}{1992}]{ppj92}
Petitjean, P., Bergeron, J., Puget, J.~L. 1992, 
\newblock {A\&A}, {265}, 375

\bibitem[\protect\astroncite{Petitjean et al.}{1994}]{ppj94}
Petitjean, P., Rauch, M., Carswell, R.~F. 1994, 
\newblock {A\&A}, {291}, 29

\bibitem[\protect\astroncite{Prochaska \& Burles}{1999}]{pb99}
Prochaska, J.~X., and Burles, S.~M. 1999,
\newblock {ApJ}, {117}, 1957

\bibitem[\protect\astroncite{Rauch et al.}{1997}]{rhs97}
Rauch, M., Haehnelt, M.~G., Steinmetz, M. 1997,
\newblock {ApJ}, {481}, 601

\bibitem[\protect\astroncite{Rauch et al.}{1999}]{rsb99}
Rauch M., Sargent W.~L.~W., Barlow T.~A. 1999, 
\newblock {ApJ}, {515}, 500


\bibitem[\protect\astroncite{Reimers et al.}{1997}]{reim97}
Reimers, D., K\"ohler, S., Wisotzki, L., et al. 1997, 
\newblock {A\&A}, {318}, 347

\bibitem[\protect\astroncite{Savage et al.}{2000}]{savage00}
Savage, B.~D., Wakker, B., Jannuzi, B.~T., et al. 2000,
\newblock {ApJS}, {129}, 563

\bibitem[\protect\astroncite{Savaglio et al.}{1997}]{sav97}
Savaglio, S., Cristiani, S., D'Odorico, S., et al. 1997, 
\newblock {A\&A}, {318}, 347

\bibitem[\protect\astroncite{Savaglio}{1998}]{sav98}
Savaglio, S. 1998,
\newblock {AJ}, {116}, 1055

\bibitem[\protect\astroncite{Savaglio et al.}{1999}]{sav99}
Savaglio S., Ferguson H.C., Brown T.M., et al. 1999,
\newblock {ApJ}, {515}, 5

\bibitem[\protect\astroncite{Schaerer \& de Koter}{1997}]{sk97}
Schaerer, D., and de Koter, A. 1997,
\newblock {A\&A}, {322}, 598


\bibitem[\protect\astroncite{Sealey et al.}{1998}]{sdw98}
Sealey, K.~M., Drinkwater, M.~J., Webb, J.~K. 1998, 
\newblock {ApJ}, {499}, L135


\bibitem[\protect\astroncite{Songaila}{1998}]{song98}
Songaila, A. 1998, 
\newblock {ApJ}, {115}, 2184

\bibitem[\protect\astroncite{Songaila \& Cowie}{1996}]{song:cow96}
Songaila, A., and Cowie, L.~L. 1996,
\newblock {AJ}, {112}, 335


\bibitem[\protect\astroncite{Steidel et al.}{1994}]{steidel94}
Steidel, C.~C., Dickinson, M., Persson, E. 1994, 
\newblock {ApJ}, {437}, L75 


\bibitem[\protect\astroncite{Sutherland \& Dopita}{1993}]{su:do93}
Sutherland, R.~S., and Dopita, M.~A. 1993, 
\newblock {ApJS}, {415}, 174

\bibitem[\protect\astroncite{Zhang et al.}{1995}]{zhang95}
Zhang, Y., Anninos, P., Norman, M.~L. 1995, 
\newblock {ApJ}, {453}, L57

\bibitem[\protect\astroncite{Zhang et al.}{1997}]{zhang97}
Zhang, Y., Anninos, P., Norman, M.~L., Meiksin, A. 1997, 
\newblock {ApJ}, {485}, 496

\bibitem[\protect\astroncite{Zheng et al.}{1998}]{zdk98}
Zheng, W., Davidsen, A.~F., Kriss, G.~A. 1998, 
\newblock {AJ}, {115}, 391
\end{thebibliography}
\bibliographystyle{apj}

\end{document}